\documentclass[12pt]{article}

\newcommand{\La}{\Lambda}
\newcommand{\prd}{\partial}

\newcommand{\nn}{ \nonumber }

\renewcommand{\arraystretch}{1.5}

\newcommand{\ice}[1]{\relax}

\newcommand{\pivo}[1]{\relax}

\newcommand{\beq}{\begin{equation}}
\newcommand{\eeq}{\end{equation}}
\newcommand{\bea}{\begin{eqnarray}}
\newcommand{\eea}{\end{eqnarray}}

\newcommand{\ba}{\begin{array}} 
\newcommand{\ea}{\end{array}} 
 
\newcommand{\als}{\alpha_s}
\newcommand{\as}{a_s}

\newcommand{\g}{\gamma}

\newcommand{\msbar}{\overline{\mbox{MS}}}

\newcommand{\ovl}{\overline}

\newcommand{\BreakII}{ \right. \right.  \nonumber \\ &{}& \left.\left. }

\newcommand{\lsb}{\left[}
\newcommand{\rsb}{\right]}

\begin{document}

\begin{titlepage}
\noindent
\hfill THEP 02/18 \\
\mbox{}
\hfill TTP02--45\\
\mbox{}
\mbox{}
\hfill hep-ph/0212303\\
\hfill {}

\vspace{0.5cm}
\begin{center}
  \begin{Large}
  \begin{bf}
Five-loop vacuum polarization in pQCD:
$\mathcal{O}(m_q^2\alpha_s^4 n_f^2)$ contribution 
  \end{bf}
  \end{Large}

  \vspace{0.8cm}

  \begin{large}
P.A.~Baikov\\
 {\small {\em 
Institute of Nuclear Physics, 
Moscow State University
Moscow~119992, Russia}}\\
K.G. Chetyrkin\footnote{On leave from Institute for Nuclear Research
of the Russian Academy of Sciences, Moscow, 117312, Russia.} \\
{\small {\em Fakult{\"a}t f{\"u}r Physik
 Albert-Ludwigs-Universit{\"a}t Freiburg,
D-79104 Freiburg, Germany }}
J.H.~K\"uhn  \\[3mm]
{\small {\em
    Institut f\"ur Theoretische Teilchenphysik,
    Universit\"at Karlsruhe,
     D-76128 Karlsruhe, Germany}
}
\end{large}

\vspace{0.8cm}
{\bf Abstract}
\end{center}
\begin{quotation}
\noindent

We present the analytical calculation of the contribution of order
$m_q^2 \alpha_s^4 n_f^2$ to the absorptive part of the vacuum
polarization function of vector currents. This term constitutes an
important gauge-invariant part of the full $\mathcal{O}(m_q^2 \als^4)$
correction to the total cross-section of $e^+ e^-$ annihilation into
hadrons.  The results are 
compared to predictions
following from various optimization schemes.

\end{quotation}
\end{titlepage}

\section{Introduction}

In QCD the correlator of two currents is a central object from which
important physical consequences can be deduced (for a detailed review
see, e.g.~\cite{phys_report}).  In particular, important physical
observables like the cross-section of $e^+ e^-$ annihilation into
hadrons and the decay rate of the $Z$ boson are related to the vector
and axial-vector current correlators.  Furthermore, total decay rates
of CP even or CP odd Higgs bosons can be obtained by considering the
scalar and pseudo-scalar current densities, respectively.

From the theoretical viewpoint the two-point correlators are
especially suited for  evaluations
in the  framework of perturbative QCD (pQCD)
\cite{Steinhauser:2002rq}. Indeed, due to the simple kinematics (only one
external momentum) even multiloop calculations can be analytically
performed. Non-perturbative contributions can be
effectively  controlled through the  the operator product
expansion \cite{svz,OPE:Wilson}.  As a consequence, the results for
practically all physically interesting correlators (vector,
axial-vector, scalar and pseudo-scalar) are available up to order
$\alpha_s^2$ taking into account the {\em full} quark mass dependence
\cite{CheKueSte96,CheHarSte98,non_diag}.

In many important cases (with the Z decay rate  a prominent
example) the external momentum  is much larger than the masses of
the (active) quarks involved. This 
justifies to neglect   these masses in a first approximation 
which significantly simplifies the calculation.
As a result, in massless QCD the vector and scalar correlators are
analytically known to $\als^3$ \cite{GorKatLar91SurSam91,gvvq,gssq}.
The residual quarks mass effects can be taken into account via the
expansion in quark masses.  At present this has been done for the
quadratic and quartic terms to the same $\als^3$ order
\cite{ChetKuhn90,mq2as3,mq4as3}.

During the past years, in particular through the analysis of $Z$
decays at LEP and of $\tau$ decays, an enormous reduction of the
experimental uncertainty (down to $\mathcal{O}(10^{-3})$) and  in $R(s)$ has been achieved
with the perspective of a further reduction by a factor of four at a
future linear collider \cite{TESLA}.

Inclusion of the $\mathcal{O}(\als^3)$ corrections
\cite{GorKatLar91SurSam91} is  mandatory already now. Quark mass
effects as well as corrections specific to the axial current
\cite{KniKue90b,CheKue93,CheTar94LarRitVer94} must be included for the case of $Z$-decays.  The
remaining theoretical uncertainty from uncalculated higher orders is
at present comparable to the experimental one
\cite{phys_report}. 
Thus, the full calculation of the next  contributions,
those of $\mathcal{O}(\als^4)$, 
to $R(s)$ is on agenda.

In   massless approximation $R(s)$ is conveniently written as
\bea
R(s) &=& \sum_{f} Q_f^2\, 3 \left( 
\rule{0mm}{7mm}
\right.
1 +
\as + \as^2  (1.98571 - 0.115295 \, n_f) + 
\nn
\\
&{}& 
\as^3  (-6.63694 - 1.20013 \, n_f - 0.00517836 \, n_f^2 ) + 
 \as^4 \,  r^{V,4}_0
\left.
\rule{0mm}{7mm}
\right)
\dots	
{},
\label{R(s):nf:as^3+rV4}
\eea
where $\as = \alpha_s(\mu^2=s)/\pi$ and the standard $\msbar$ renormalization
prescription \cite{msbar} is understood.
The $\as^4$ term can be further decomposed as 
a polynomial in $\, n_f$, namely
\beq
r^{V,4}_0 = r^{V,4}_{0,0} + r^{V,4}_{0,1} \, n_f + r^{V,4}_{0,2} \, n_f^2
+ r^{V,4}_{0,3} \, n_f^3 
{}.
\label{rV:nf expansion}
\eeq
The term of order $\als^4 \, n_f^3$ (and, in fact, all terms of order
$\als(\als \, n_f)^n$) have been obtained earlier by summing the
renormalon chains \cite{VV:renormalons}. These are  technically very simple
to compute\footnote{At least for a fixed number of loops.}, 
numerically small ($r^{V,4}_{0,3} = 0.02152$) and not directly sensitive to the nonabelian
character of QCD, in contrast to the terms of order $\als^4 n_f^2$. 
The first nontrivial contribution --- 
the subleading ${\mathcal{O}}(n_f^2)$  piece in (\ref{rV:nf expansion}) ---
have been only recently analytically  computed in \cite{Baikov:2001aa} with the 
result\footnote{For brevity we display it in the numerical form.}
\beq
r^{V,4}_{0,2} = -0.7974
{}.
\label{rV4nf2:result}
\eeq

Taken by itself eq.~(\ref{rV4nf2:result}) is  not of much use for
the phenomenology for obvious reasons. However, 
it gives a  strong  extra support to the well-known  prediction
of the full ${\mathcal{O}}(\alpha_s^4)$ contribution  
of the Kataev and Starshenko\cite{KS} (for a detailed discussion see 
\cite{BKCh:tau}).
The prediction is often used to estimate the theoretical uncertainty 
in $R(s)$ due to higher order not yet computed  perturbative
contributions.  

Another point of concern are the quark mass effects at order
$\alpha_s^4$. Fortunately, simple estimates (see
e.g. \cite{ChetKuhn90}) show that, say, the quadratic mass effects are
completely under control\footnote{ \emph{Provided} one uses the
properly chosen parameterization in terms of the running quark mass.}
at the scale of the Z-boson mass since terms of order
${\mathcal{O}}(\alpha_s^3)$ are available and small. This holds true
for the vector correlator with the leading term proportional 
$\alpha_s m_b^2$ as well as for the axial vector correlator, where the the
leading in $\alpha_s$ term is present already at the Born level and
thus proportional to $m_b^2$.  Quarks mass effect are getting
progressively more important at lower energies.  From the conceptual
viewpoint it
 would be  important  to test the evolution of the
strong coupling as predicted by the beta function and the 
standard QCD matching procedure through a
determination of $\as$ from essentially the same observable, however,
at lower energy. The region from several GeV above charm threshold
(corresponding to the maximal energy of BEPC around $5.0$ GeV) to just
below the $B$ meson threshold at around $10.5$ GeV corresponding to
the ``off resonance" measurements of CESR or  B-meson factory is  particularly suited
for this purpose.  As a consequence of the favorable error
propagation, 
\[
\delta \as(s) = \frac{\as^2(s)}{\as^2(M_Z^2)} \delta \as(M_Z^2)
{},
\]
the accuracy in the measurement (compared to $91$ GeV)
may decrease by factor of about 3 or even 4 at 10 and 5.6 GeV
respectively, to achieve comparable precision in 
$\La_{QCD}$.

Technically speaking, the evaluation of the leading in $n_f$ term of
order $O(m_q^2\alpha_s^4 n_f^3)$ is again rather simple while the
subleading term of order $O(m_q^2\alpha_s^4 n_f^2)$ presents a problem
comparable to that in the massless limit.  In this work we describe
the corresponding calculation and its results.  We limit the
discussion to the vector current correlator, which is relevant for
hadron production through the electromagnetic current. We also compare
our results with predictions following from various optimization
schemes.  Our basic conclusion is once again that the FAC/PMS
optimization predictions are remarkably close to reality. They
provides a quantitative argument in favour of the corresponding full
prediction.

\section{Generalities}
To fix notation we start  considering  two-point correlator
of vector  currents and the corresponding vacuum polarization
function  ($j_\mu^v= \ovl{Q}\gamma_\mu Q$;
$Q$ is a  quark field with  mass $m$, all other $n_f-1$ quarks  are
assumed to be massless)
\begin{eqnarray}
\Pi_{\mu\nu}(q)  &=&
 i \int {\rm d} x e^{iqx}
\langle 0|T[ \;
\;j_{\mu}^{v}(x)j_{\nu}^{v}(0)\;]|0 \rangle
=
\displaystyle
(-g_{\mu\nu}q^2  + q_{\mu}q_{\nu} )\Pi(q^2)
{}.
\label{PiV}
\end{eqnarray}
The  physical observable $R(s)$ is related to $\Pi(q^2)$
by
\beq
R(s) = 12 \pi \Im \, \Pi(q^2 + i\epsilon)
{}.
\label{R(s):def}
\eeq
It is convenient to decompose $R(s)$ into the  massless and the quadratic
terms
\bea
R(s) &=& 3 \left\{r^V_0 +  \frac{m^2}{s} r^{V}_2 \right\}+\dots 
= 3\left\{\sum_{i \ge 0} a_s^i\left(
 r^{V,i}_0 + \frac{m^2}{s} r^{V,i}_2 \right)
                                    \right\} +\dots
\nn
{}.
\eea
Here we have set the normalization scale  $\mu^2= s$;
$a_s = \als(s)/\pi$ and dots stand for corrections proportional
to $m^4$ and higher powers of the quark mass.

For the calculation of $r^{V,4}_0$ we  had to deal with divergent
parts of five-loop diagrams and finite parts of the four-loop ones
(see, e.g the corresponding discussion in
\cite{gvvq,Chetyrkin:uw}). 
In fact, as was first discovered in
Ref.~\cite{ChetKuhn90}, the quadratic mass correction $r^{V,4}_2$ can
be obtained with the help of renormalization group methods {\em
exclusively} from the four-loop function $\Pi_2$ defined as
($ a_s = \als(\mu^2)$ )
\[
\nn
\Pi   = 
\frac{3}{16 \pi^2} \left(
\Pi_0  + \frac{m^2}{-q^2} \Pi_2
\right)
{},
\ \ \ 
\Pi_2 = \sum_{i \ge 0}\   a_s^i k^{V,i}_2
{}.
\]
Indeed, due to an extra factor of $m^2$ the function $\Pi_2$  obeys an
\emph{unsubtracted} dispersion relation. As a result the following
\emph{homogeneous} RG equation  holds:
\beq
\left(
\mu^2\frac{\partial}{\partial\mu^2}
 +
\g_m(\as) m \frac{\partial}{\partial m}
 +
\beta(a_s)
a_s
\frac{\partial}{\partial a_s}
\right)
      \Pi_2
          = 0
\label{rg:Pi2}
{}.
\eeq
Equivalently, since $\Pi_2$ is independent of $m$, 
\beq
\frac{\prd }{\prd L} \Pi_2 =
\left(
-2\g_m(\as)
-
 \beta( a_s) a_s\frac{\prd }{\prd a_s}
\right)
\Pi_2
\label{rg:Pi2:L}
{},
\eeq
where  $L = \ln\frac{\mu^2}{-q^2}$.
Once the rhs of eq.~(\ref{rg:Pi2:L}) is  known, one can  find  the function 
$\Pi_2$ up to a constant contribution which  has no effect   on $R(s)$.  
Thus, to compute, say, 
the  $\mathcal{O}(\frac{m^2}{s} \as^3)$ correction to $R(s)$ one  needs to know
\begin{itemize}
\item
The three-loop QCD $\beta$ function
function  (starting at ${\mathcal{O}} (\alpha_s^2)$) and 
the quark mass anomalous dimension (starting at ${\mathcal{O}}(\alpha_s)$):
available from    \cite{beta3loop,gamma3loop}, each 
up to order  $\alpha_s^3$.
\item
The very polarization operator  $\Pi_2$ at {\em three} loops
(that is  of order  $\alpha_s^2$):
was computed also long ago in \cite{GorKatLar86}.
\end{itemize}
Altogether, this implies that the ${\cal O}(\frac{m^2}{s} \as^3)$  term in $R(s)$ 
\emph{could} have been been computed as early as in 1986. But in reality this  
was done only by 5 years later \cite{mq2as3}.

From the above discussion it is clear which ingredients are
necessary to compute quadratic  mass terms in $R(s)$ at ${\cal O}(\as^4)$:
\begin{enumerate}
\item
The four -loop QCD $\beta$ function
function and quark mass anomalous dimension:
available  from \cite{RitVerLar97,Che97,LarRitVer972}.
\item
The  polarization operator $\Pi_2$ at {\emph four} loops,
including its  {\em constant} part.
\end{enumerate}

\section{Results}

Using the new technique described in \cite{bai1,bai2,Baikov:2001aa}
and the parallel version of FORM \cite{Fliegner:2000uy,Vermaseren},
we have computed the leading and subleading (in $n_f$) contributions of order 
$\als^3$   to the polarization function $\Pi_2$
($\mathcal{O}(\alpha_s^2)$ results are known from \cite{GorKatLar86})
\bea
 \Pi_{2} = -8 &-&\frac{64}{3}\, a_s + a_s^2 \left\{
\frac{95}{9} \,n_f +-\frac{18923}{54}-\frac{784}{27} \,\zeta_{3}+\frac{4180}{
27} \,\zeta_{5}\right\}
\nonumber
 \\ &+& a_s^3 \left\{
 \lsb -\frac{5161}{1458}-\frac{8}{27} \,\zeta_{3} \rsb  \, n_f^2 +\lsb \frac{
62893}{162}+\frac{424}{27} \,\zeta_3^2-\frac{4150}{243} \,\zeta_{3}\BreakII
 \phantom{ a_s^3 ||}
+\frac{20}{3} \,\zeta_{4}-\frac{28880}{243} \,\zeta_{5} \rsb  \,n_f +k^{[V]3}
_{2,0}\right\} 
{}
\label{Pi2:nf} \end{eqnarray}
or, numerically,
\beq
\Pi_2 =
-  8 - 21.333 \, a_s 
+  a_s^2 \, ( 10.56 \, n_f -224.80 ) 
+ a_s^3\, (- 3.896 \, n_f^2 +  274.37\, n_f   + k^{V,3}_{2,0} )
{},
\label{Pi_2:num}
\eeq
where we have  set  $a_s = \alpha_s(-q^2)/\pi, \mu^2 = -q^2$. 

The  straightforward  use of eqs.~(\ref{R(s):def},\ref{rg:Pi2:L}) yields 
($a_s = \alpha_s(s)/\pi, \mu^2 = s$)
\bea r^{V}_{2} = 12 \frac{\alpha_s}{\pi} &+& a_s^2 \left\{
-\frac{13}{3} \,n_f +\frac{253}{2}\right\}
\nonumber
 \\ &+& a_s^3 \left\{
 \lsb \frac{125}{54}-\frac{1}{9} \pi^2 \rsb  \, n_f^2 +\lsb -\frac{4846}{27}+
\frac{17}{3} \pi^2-\frac{466}{27} \,\zeta_{3}\BreakII
 \phantom{ a_s^3 ||}
+\frac{1045}{27} \,\zeta_{5} \rsb  \,n_f +2442-\frac{285}{4} \pi^2+\frac{490}
{3} \,\zeta_{3}-\frac{5225}{6} \,\zeta_{5}\right\}
\nonumber
 \\ &+&a_s^4\left\{
 \lsb -\frac{2705}{1944}+\frac{13}{108} \pi^2 \rsb  \, n_f^3 +\lsb \frac{9194
3}{486}-\frac{1121}{108} \pi^2+\frac{53}{9} \,\zeta_3^2\BreakII
 \phantom{a_s^4||}
+\frac{13}{324} \,\zeta_{3}-\frac{3610}{81} \,\zeta_{5} \rsb  \, n_f^2 + r_{2
,0}^{V,4} \,n_f +r_{2,1}^{V,4}\right\} {}.
\label{rV4_2V:nf} \end{eqnarray}
Numerically
\bea
r^V_2 = {}
12\,  a_s &+&   a_s^2 \left(-4.3333 \ n_f+  126.5  \right)
+
a_s^3
\left(
 1.2182 \, n_f^2- 104.167  \, n_f  + 1032.14 
\right)
\nn
\\
&+& a_s^4
\left(
- 0.20345 \,  n_f^3
+ 49.0839  \,n_f^2  
+ \  r^{V,4}_{2,1}\, n_f \ + r^{V,4}_{2,0}
\right)
{}.
\label{rV_2_2:num}
\eea

\section{Discussion}
Let us contrast eq.~(\ref{Pi2:nf}) with predictions based on the
optimization methods FAC (Fastest Apparent Convergence) and PMS
(Principle of Minimal Sensitivity)
\cite{ste,gru,krasn,KS}. As is well known in many cases FAC and PMS
predictions are quite close \cite{Kataev:1994wd}. To compare with an
independent  
prediction   
we will also use so-called NNA (Naive
NonAbelianization) approach of \cite{BG}.

In Table 1 we list the FAC and PMS predictions for the 
function $\Pi_2$ 
(see $\leftarrow$ \cite{CheKniSir97}). 
{}From the three entries corresponding to $n_f = 3,4$ and $5$ one
easily restores the FAC/PMS prediction for the $n_f$ dependence of the
$\as^3$ term in $\Pi_{2}$ 
\bea 
k^{V,3}_{2}(\mathrm{FAC}) &=& -4.2 \, n_f^2 + 277 \, n_f -2886    
\label{K4:FAC:nf}
{},
\\
k^{V,3}_{2}(\mathrm{PMS}) &=& -4.2 \, n_f^2   + 282 \, n_f   -2916
\label{K4:PMS:nf}
{}.
\eea
The comparison with  (\ref{Pi_2:num}) shows  very good agreement with
the calculated  terms of order $n_f$ and $n_f^2$.  As a natural next
step we will now use the available \emph{exact} information about the
coefficients $k^{V,3}_{2,2}$ and $k^{V,3}_{2,1}$ to find
$k^{V,3}_{2,0}$ from FAC/PMS prediction for a selected value of $n_f=
3$. The choice seems to be natural as in many cases FAC/PMS
predictions made for $n_f= 3$ are in better agreement with the exact
results (see. e.g. \cite{KS,Kataev:1994wd,BKCh:tau}).  We
obtain (in fact, very close values are produced with $n_f =4$ and 5)
\beq
k^{V,3}_{2,0}(\mathrm{FAC},nf = 3) =  -2880, \   \ \ k^{V,3}_{2,0}(\mathrm{PMS},nf = 3) =  -2897
{}.
\label{FAC/PMS:nf:3}
\eeq

Now we turn to NNA. The corresponding prescription is to simply
make the substitution  $n_f \to -3/2\beta_0 = n_f -33/2$ in  the 
leading term of the $n_f$ expansion. The result is
\beq 
k^{V,3}_{2}(\mathrm{NNA})  =  -1060.69  + 128.568 \, n_f  - 3.896 \, n_f^2  
{}.
\label{NNA}
\eeq
This  prediction is in obvious conflict  with  the
result of  the direct calculation of the term linear in $n_f$.

\begin{table}[hbt]
\caption{\label{t1}
Estimates for  the  coefficient $k^{V,3}_2$
based on  FAC and PMS optimization schemes.
}
\renewcommand{\arraystretch}{1.2}
\begin{center}
\begin{tabular}{|c|c|c|c|}
\hline
Method& $n_f=3$ & $n_f=4$ & $n_f=5$  \\
\hline\hline
FAC & $-2092  $  & $ - 1844 $ & $- 1604 $ \\
\hline
PMS & $ -2109 $ & $-1856$ & $-1612 $\\
\hline
\end{tabular}
\end{center}
\end{table}

\section{Conclusions}

We have presented the analytical calculation of contributions of
order $m_q^2 \alpha_s^4 n_f^2$ to the vacuum polarization function of
vector currents. Predictions of FAC, PMS- and NNA optimization methods
are tested against the exact results.  Good quantitative agreement
is found for the first two cases.  NNA  
seemingly  predicts only sign and  order of magnitude
of analytical results.

The authors are grateful to M.~Steinhauser   
and R.~Harlander for the careful reading 
of the manuscript and   very useful comments.  This work was supported by the
DFG-Forschergruppe {\it ``Quantenfeldtheorie, Computeralgebra und
Monte-Carlo-Simulation''} (contract FOR 264/2-1), by INTAS (grant
00-00313), by RFBR (grant 01-02-16171), by Volkswagen Foundation and
by the European Union under contract HPRN-CT-2000-00149.

\end{document}